\documentclass[%
 superscriptaddress,
 aip,
 amsmath,amssymb,
preprint,%
 author-year,%
longbibliography
]{revtex4-2}

\usepackage{amsmath}
\usepackage{graphicx}
\usepackage{textcomp}

\usepackage{hyperref, color}
\definecolor{linkColor}{rgb}{0.8,0,0}
\definecolor{darkred}{rgb}{0.8,0,0}
\hypersetup{pdfborder={0 0 0},colorlinks=true,urlcolor=linkColor,citecolor=linkColor}

\usepackage{fancyhdr}
\begin{document}

\title{Evaluating Stage Motion for Automated Electron Microscopy}

\author{Kevin R. Fiedler}
\affiliation{College of Arts and Sciences, Washington State University -- Tri-Cities, Richland, Washington 99354}
\affiliation{Energy and Environment Directorate, Pacific Northwest National Laboratory, Richland, Washington 99352}

\author{Matthew Olszta}
\affiliation{Energy and Environment Directorate, Pacific Northwest National Laboratory, Richland, Washington 99352}

\author{Kayla Yano}
\affiliation{Energy and Environment Directorate, Pacific Northwest National Laboratory, Richland, Washington 99352}

\author{Christina Doty}
\affiliation{National Security Directorate, Pacific Northwest National Laboratory, Richland, Washington 99352}

\author{Derek Hopkins}
\affiliation{Environmental Molecular Sciences Laboratory, Pacific Northwest National Laboratory, Richland, Washington 99352}

\author{Sarah Akers}
\affiliation{National Security Directorate, Pacific Northwest National Laboratory, Richland, Washington 99352}

\author{Steven R. Spurgeon}
\email{steven.spurgeon@pnnl.gov}
\affiliation{Energy and Environment Directorate, Pacific Northwest National Laboratory, Richland, Washington 99352}
\affiliation{Department of Physics, University of Washington, Seattle, Washington 98195}

\date{\today}


\begin{abstract}

Precise control is an essential and elusive quality of emerging self-driving microscopes. It is widely understood these instruments must be capable of performing rapid, high-volume, and arbitrary movements for practical self-driving operation. However, stage movements are difficult to automate at scale, owing to mechanical instability, hysteresis, and thermal drift. Such difficulties pose major barriers to intelligent microscope designs that require repeatable, precise movements. To guide design of emerging instruments, it is necessary to understand the behavior of existing designs to identify rate limiting steps for full autonomy. Here we describe a general framework to evaluate stage motion in any electron microscope. We define metrics to evaluate stage degrees of freedom, propose solutions to improve performance, and comment on fundamental limits to automated experimentation using present hardware.

\end{abstract}

\maketitle

\section{Introduction}

Artificial intelligence (AI), encompassing disciplines such as robotics, machine learning (ML), and computer vision, has begun to transform the study of materials, chemical, and biological systems \citep{Batra2021, Sha2020, Lalmuanawma2020}. AI allows us to richly analyze more data and discover latent, multidimensional patterns that inform physical mechanisms, such as those underpinning quantum computing, energy storage, and designer medicine \citep{Schmidt2019, Vasudevan2019, Butler2018, Battineni2020}. Electron microscopy (EM), a pillar of characterization at high spatial and chemical resolution, stands to benefit greatly from these approaches\citep{Kalinin2022, Treder2022, Ede2020}. At present, most EM data collection and analysis is still conducted by hand, with limited subsets of data collected from large or rapidly changing samples. With the emergence of customized ML techniques and inexpensive edge computing hardware, it is now possible to analyze data in greater volume and depth. ML methods have shown success on a range of EM-related tasks, including segmentation,\citep{Stuckner2022, Akers2021, Groschner2021, Xu2021} automated instrument tuning, \citep{Xu2022} determination of microstructural descriptors, \citep{Dan2022, Ziatdinov2022, Ziatdinov2017, Laanait2016} and in situ forecasting \citep{Lewis2022}. These methods are now beginning to grapple with the large volumes of data produced by modern detectors, yielding richer statistical insights into important chemical and materials systems \citep{Spurgeon2020c}.

While new approaches for EM data analysis have been developed, their on-the-fly implementation on microscope hardware has been far slower. This situation is largely due to to proprietary, vendor-defined platforms that limit direct instrument control, obscure underlying raw data, and execute ``black box'' analysis routines, all of which are fundamental barriers to findable, accessible, interoperable, and reusable (FAIR) microscopy \citep{Kalinin2021a, Schorb2019, Wilkinson2016}. Additionally, these ``black box'' routines often hinder reproducibility and high-quality microscopy, given that the underlying methodologies are often misunderstood by standard users. Recently, there have been some notable successes in the development of more open microscope platforms and controllers, including systems driven by Gaussian process optimization, \citep{Liu2022, Ziatdinov2022a} and our own platform based on sparse data analytics \citep{Olszta2021}. In all cases, numerous decisions must be made regarding imaging conditions, sample movement/orientation, and detector configuration dictated by sample analysis requirements. In the most basic ``open-loop'' experiment these decisions are hard-coded a priori and then executed without further adaptive feedback. While there are some scenarios in which an ``open-loop'' approach can be successful, such as large area, low magnification imaging of static samples, the approach is unsuitable for most high-resolution imaging or in situ experimentation. To truly be useful in common experiments, a more powerful ``closed-loop'' approach is required. In this approach, the microscope system performs on-the-fly human-like reasoning to detect changes relative to control set points, such as the movement to a region of interest or changes in a spectrum. This first step requires domain-specific analytics, which has been the topic of many recent studies \citep{Ziatdinov2022, Ghosh2022, Treder2022, Stuckner2022, Akers2021, Muto2020}. However, once an instrument control decision is made it must be implemented precisely. Here vendor-locked systems pose major barriers, since it is often not possible to directly program the microscope, though sophisticated workarounds have been developed by the community \citep{Yin2020, Mastronarde2003a, Carragher2000}. New microscope controllers are beginning to emerge that take advantage of low-level application programming interfaces (APIs), open-source software, and fast computing hardware to direct decision-making \citep{Olszta2021, Roccapriore2022}.

With the advent of such new controllers, the community must now address the challenge of implementing a control platform that can account for both large volumes of decisions and potential imprecision in their execution. We have recently developed a centralized, asynchronous scanning transmission electron microscope (STEM) control platform that has allowed us to evaluate challenges to self-driving ``closed-loop'' microscopy \citep{Olszta2021}. As already mentioned, there are many sources of imprecision in microscope control; for example, lens voltages can drift, leading to parasitic aberrations, or mechanical lash can cause hysteresis in stage movements. In practice, this means that 1000s of decisions can be made in the span of a few seconds and any error compounds rapidly. It is important to note that in any self-driving experiment, errors in position will grow linearly with the number of movement commands. This is then compounded five-fold due to the number of degrees of freedom that are present with an electron microscope ($x$, $y$, $z$, $\alpha$, and $\beta$). Even a 1\% error in displacement will translate to being off by a full movement command after 100 steps, unless the motion is actively compensated. Traditionally, human operators would perform learned corrections to these errors as reflexive memory commensurate with years of experience. However, in truly automated microscopy this imprecision must be accounted for algorithmically.

To aid in the development of ``closed loop'' control we must first consider the microscope stage, which is essential for reliable imaging and observation of objects along desired orientations. There are several key characteristics of any stage: freedom of movement, stability, and movement reproducibility. Freedom of movement is largely defined by the geometry of the microscope pole piece and the capabilities of the stage itself (i.e., tilt axes, in situ stimuli, X-ray background). The static and dynamic stability of the stage are critical factors in both extreme atomic-resolution imaging, as well as during in situ experimentation, where an object must often be tracked or kept within a field of view \citep{Zheng2015}. There are many factors that contribute to stability, including the holder geometry (side entry vs. internal stage), sample response, and column temperature, and vendors have spent a great deal of time optimizing stage stability. Unfortunately, the last characteristic, movement reproducibility, has largely been overlooked in the past decades, despite the fact that it is one of the most critical parts of emerging self-driving experimentation. The ability to precisely move to and recall positions is critical for quantitative mapping, adaptive sampling, predictive tilting, and many other desirable experiments. Present automation software must perform tedious and time-consuming iterative correction to compensate for imprecision at even low magnification, or a human must be present to manually reposition the stage when an object leaves the field of view. Such correction is increasingly unfeasible if we are to move toward more automated and eventually autonomous microscopy.

Here we provide a mathematical framework to evaluate the precision and accuracy of stage movement in any electron microscope. Our aim is to provide a systematic approach to evaluate stage performance and identify barriers to fully automated experimentation. We consider the unique characteristics of scanning electron microscope (SEM) and TEM stages and define appropriate coordinate frames of reference. We then perform systematic automated tests in the TEM specifically, utilizing our AutoEM system to conduct large-scale data collection. We identify sources of error and comment on the considerations for a hypothetical future microscope stage, which will enable unprecedented new scientific discoveries.

\section{Results and Discussion}

\begin{figure}[h!]
\includegraphics[width=\textwidth]{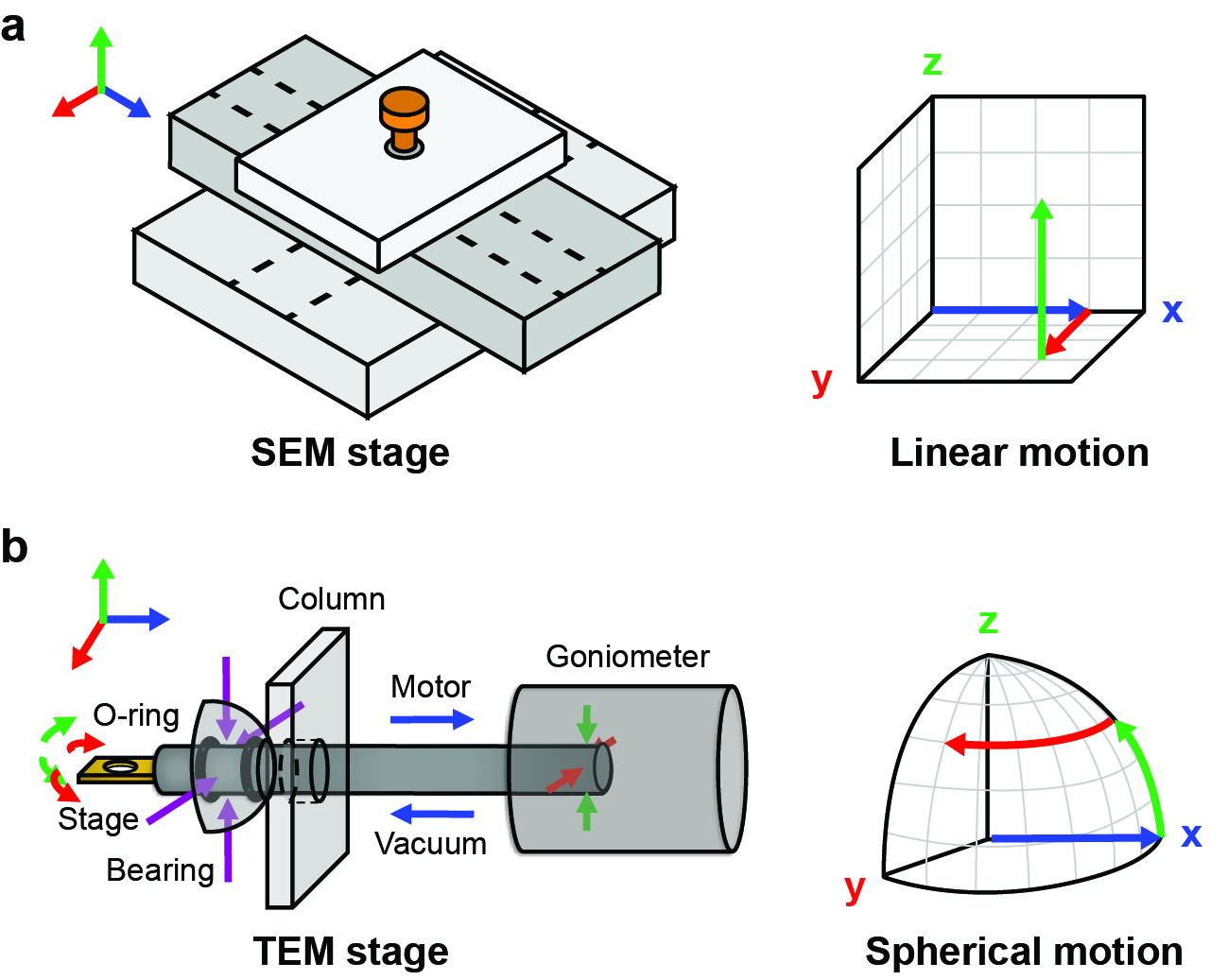}
\caption{\textbf{Overview of electron microscope stages.} (a) Simplified illustration of SEM stage and its associated linear frame of reference. (b) Simplified illustration of TEM stage and its associated spherical frame of reference. \label{stage_design}}
\end{figure}

The ability to accurately predict and control stage motion in the TEM is paramount to achieving automated experiments, such as montaging and tilt series. To understand present challenges, we must consider typical hardware designs, shown in Figure \ref{stage_design}. In the case of the SEM  (Figure \ref{stage_design}.a), movement occurs within a Cartesian frame of reference. Considering only translation, these stages generally travel linearly in the $x$ (red arrow) and $y$ (blue arrow) planes, usually along independent rails, with $z$ height (or working distance) achieved by the stage moving linearly in the $z$ direction (green arrow). In contrast, movement of a TEM stage is often confusing because either the end of the holder is encased in a protective covering (in the case of modern microscopes) or only the end of the holder is visible. We have found that users typically assume stage movement in the TEM is linear, as is the case for the SEM, since movement is only observed through cameras or a viewing screen and controls provide linear designations.

TEM stages in fact do not move in a linear fashion, but operate within a spherical coordinate system, as shown in Figure \ref{stage_design}.b. The $x$ axis can be considered largely linear, since the sample rod is pushed out by a motor (blue arrow) and then the vacuum pulls the holder to achieve the opposite direction, similar to the linear motion in the SEM. However, the complete motion of the TEM stage is occurs within a spherical frame of reference, because the goniometer pushes on the ends of the holder in both the $y$ (red arrow) and $z$ (green arrow) directions. The holder then pivots on an internal hemispherical bearing just inside the column, creating sample motion on a spherical coordinate system. The combination of these three axes makes the prediction of stage movement more difficult, with microscope manufacturers typically having already calibrated stage motion to appear to travel in a linear manner. We note that, while not shown in the diagram, the motion is further complicated by the addition of $\alpha$ (around the long axis) and $\beta$ (across the long axis) tilts. When considerations are made for the programming of stage motion, it is important to understand and quantify error associated with such spherical motion.

\clearpage

\subsection{Mathematical Framework}

We will now focus on defining a mathematical framework to evaluate stage movement in the TEM specifically. We will use the terms \textbf{target stage position}, \textbf{reported stage position}, and \textbf{actual stage position} as vernacular. The \textbf{target position} is the desired location, or input, as passed from the user to the microscope. The \textbf{reported position} is the location reported by the microscope (note this can be further considered as the displayed value or ``program'' value) after a move command has been completed, and the \textbf{actual position} is where the sample is located in space as measured by relative displacements between images using cross correlation image processing techniques, as described in Section \ref{methods}.

\begin{figure}
\includegraphics[width=\textwidth]{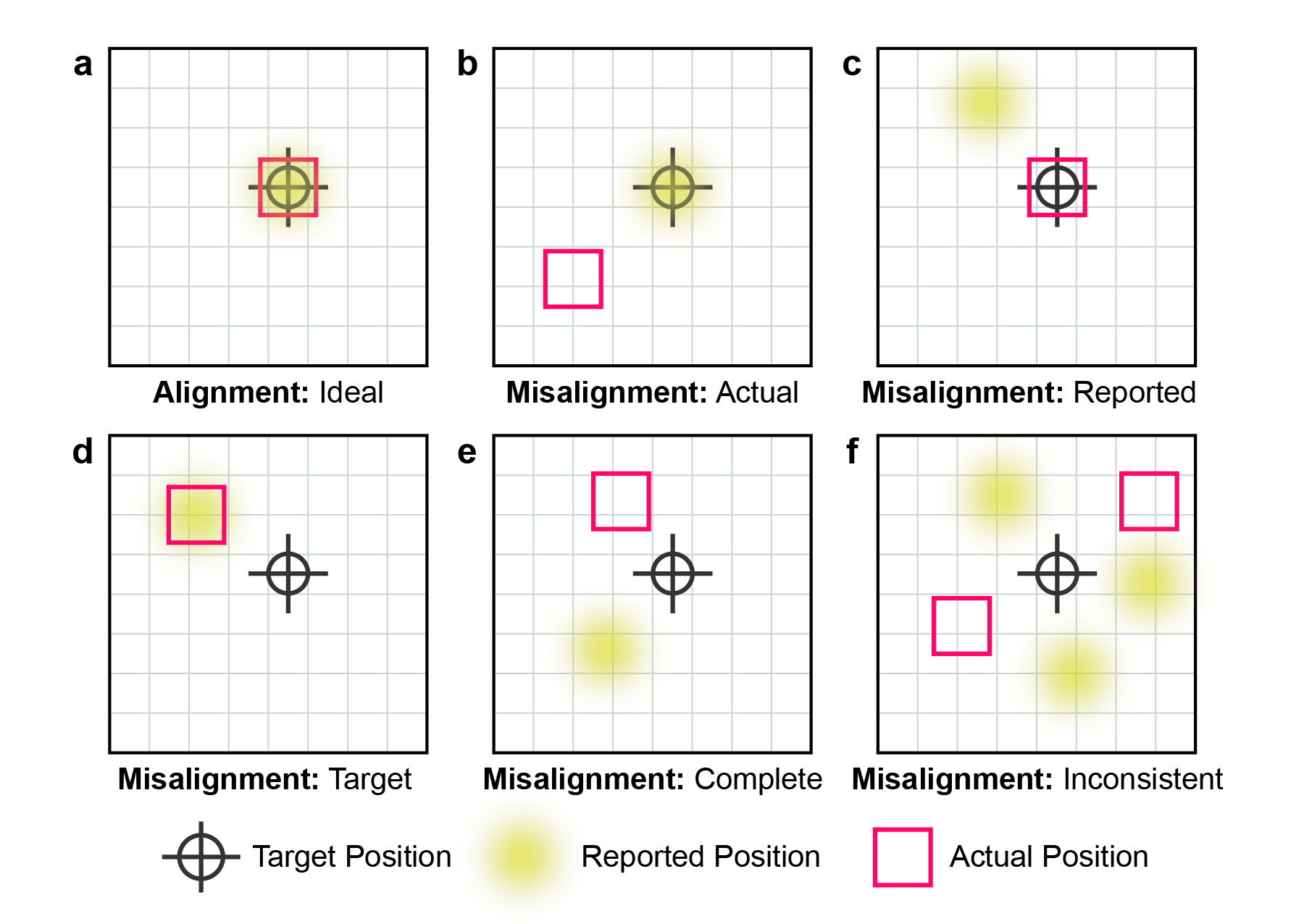}
\caption{\textbf{Framework to analyze possible microscope alignment scenarios.} (a) Target, reported, and actual positions all coincide. (b) Target and reported positions align, actual position is somewhere else. (c) Target and actual positions align, reported position is somewhere else. (d) Reported and actual positions align, target position is somewhere else. (e) None of the target, reported, or actual positions align. (f) Target, reported, or actual positions may or may not align in an inconsistent manner. The positions show either large statistical errors or state dependence, such as hysteresis. \label{alignment_cases}}
\end{figure}

To illustrate the possible interactions between these three, several cases are displayed in Figure \ref{alignment_cases}. The first case, shown in Figure \ref{alignment_cases}.a, is the ideal case for alignment, where all three terms are commensurate as desired by the user: the target, reported, and actual positions all agree. The remainder of Figure \ref{alignment_cases} illustrates cases that would be considered misalignment and are representative of most modern goniometer/stage control. These misalignments are divided into actual, reported, and target, where within various combinations of the three, two of the three positions are aligned with the third being different (Figures \ref{alignment_cases}.b--d). The next example, shown in Figure \ref{alignment_cases}.e, illustrates the case when none of the positions align with each other. While all but the first of these situations are not ideal, if the behavior is repeatable and measurable, then the user can compensate for this misalignment to achieve better performance using predictive algorithms.  This compensation is the major differentiator; the final case, Figure \ref{alignment_cases}.f, is a situation where it is not possible to compensate for the misalignment. This situation could be due to a time or state dependency, but is most likely a function of the physical state of the holder o-ring due to a degree of practical longevity.

Using this nomenclature, the ideal alignment case (Figure \ref{alignment_cases}.a) is the easiest situation where everything works as a user would expect. The next four (Figures \ref{alignment_cases}.b--f) can be compensated for, and many times are transparent to a user who is manually aligning features of interest. Since a regular user will only interact with the movement relative to the reported value, they can target a region of interest using experience and intuition. However, these situations pose a difficulty if one would like to recall a feature or perform automated, repeatable runs. The final case is the worst, because external analysis cannot compensate or improve microscope behavior. With this vocabulary, the misalignment cases would be considered forms of systematic error on the measurement. This measurement is distinguished from statistical error relating to noise in the reported value or step size. Another way of describing this would be to say that the ideal alignment has accuracy (measured close to the true value), as does Figure \ref{alignment_cases}.c. Colloquially, the other cases could be described as inaccurate measurements of the actual position.

\subsection{Hardware Testing}

 Having defined these microscope/stage input parameters, as well as the combinatorial possibilities of all three positions, we conducted a set of automated experiments to understand and quantify stage reliability/reproducibility. This approach allows us to assess the feasibility of programming autonomous modules and also enables comparisons against stages across vintage, institution, company, or microscope. We specifically outline an analysis framework where a user can understand which case their microscope exhibits and what can be done to compensate for any misalignment that is present. Generally, the idea is to complete pairwise comparisons between different position measurements (target vs. reported, target vs. actual, reported vs. actual). These should be evaluated with respect to both the consistency of measurements and whether the stage exhibits state dependence. It is important to note that this can be applied to any electron microscope stage and that the goal of the present study is not to perform a systematic evaluation of trends in microscope stages. A more comprehensive list of experimental considerations can be found in the Supplementary Section 1 and more testing results are in Supplementary Section 2.

\begin{figure}
\includegraphics[width=\textwidth]{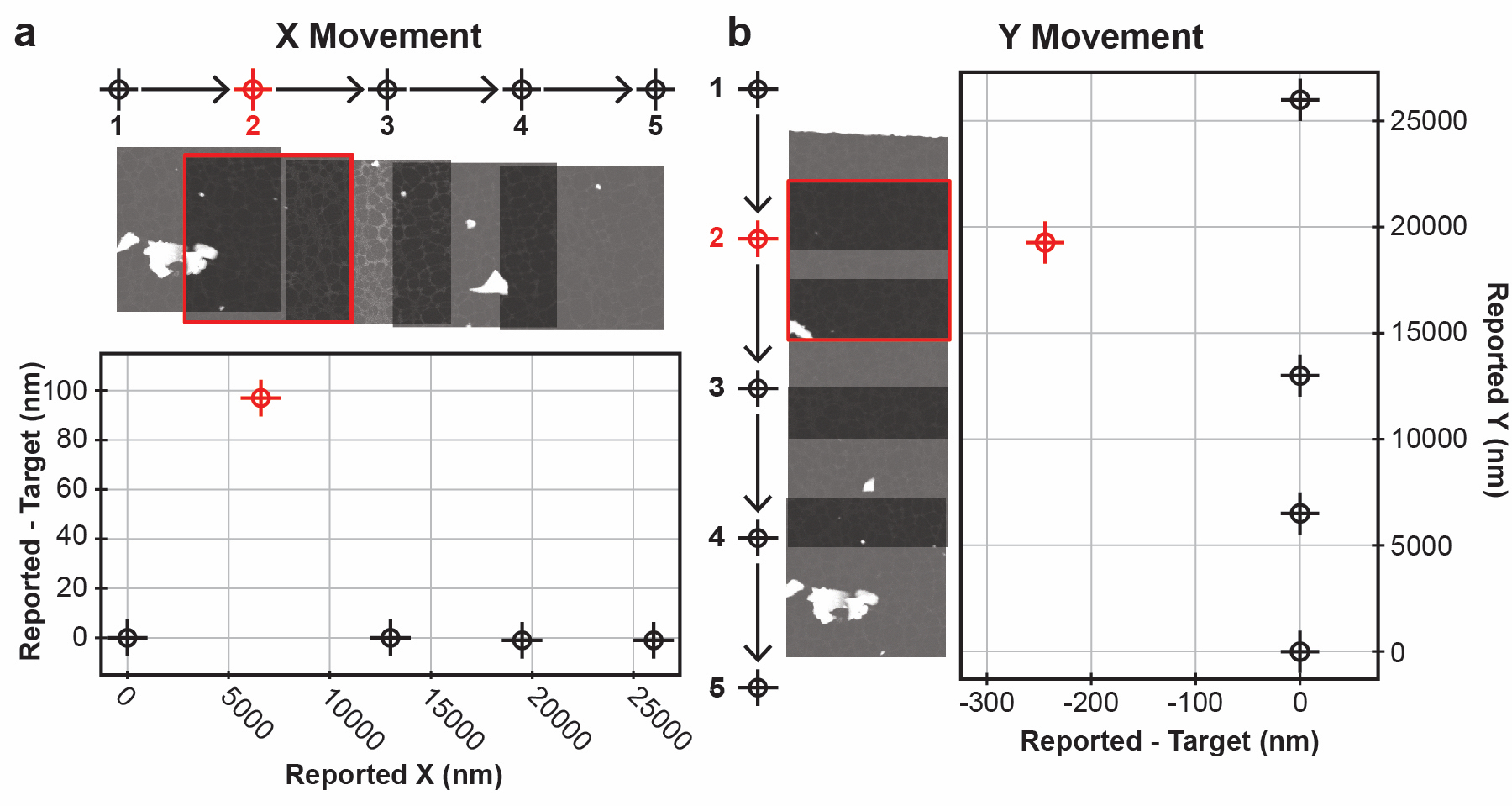}
\caption{\textbf{Hysteresis testing for independent axes.} (a) Hysteresis test for movement in the $x$ direction. (b) Hysteresis test for movement in the $y$ direction. Images were acquired sequentially along a line, and the reported position minus the target position is shown as a function of the reported position for each direction. Each image was acquired at $50,000\times$ magnification with a field of view of 3.75 \textmu m by 3.75 \textmu m and has been made 50\% transparent to illustrate overlaps. \label{hysteresis}} 
\end{figure}

To check for hysteresis on an axis, the microscope must make repeated moves in a direction and check to see if all the step sizes are the same. For the $x$ and $y$ directions, one can see the results of comparing the target and reported positions, as shown in Figure \ref{hysteresis}. For each direction, a sequence of five steps was taken. At each location an image was acquired and the reported position of the microscope noted; then a movement command was issued to move to the next position. In this figure, the difference between the reported and target position is plotted against the reported value for that axis. Generally, the agreement is fairly good, but the first step in a given direction shows a smaller reported displacement than the target. These steps are highlighted in red, and in the overlapped images, these show significantly more overlap than subsequent steps. The stitched images show a characteristic hysteretic behavior that is often referred to as backlash. In practice, backlash is accounted for by overcompensating in the reverse direction by the human operator, which must be explicitly programmed for automated experimentation. If not accounted for, objects can be incorrectly positioned or missed and these errors will accumulate over the course of an automated experiment.

\begin{figure}
\includegraphics[width=\textwidth]{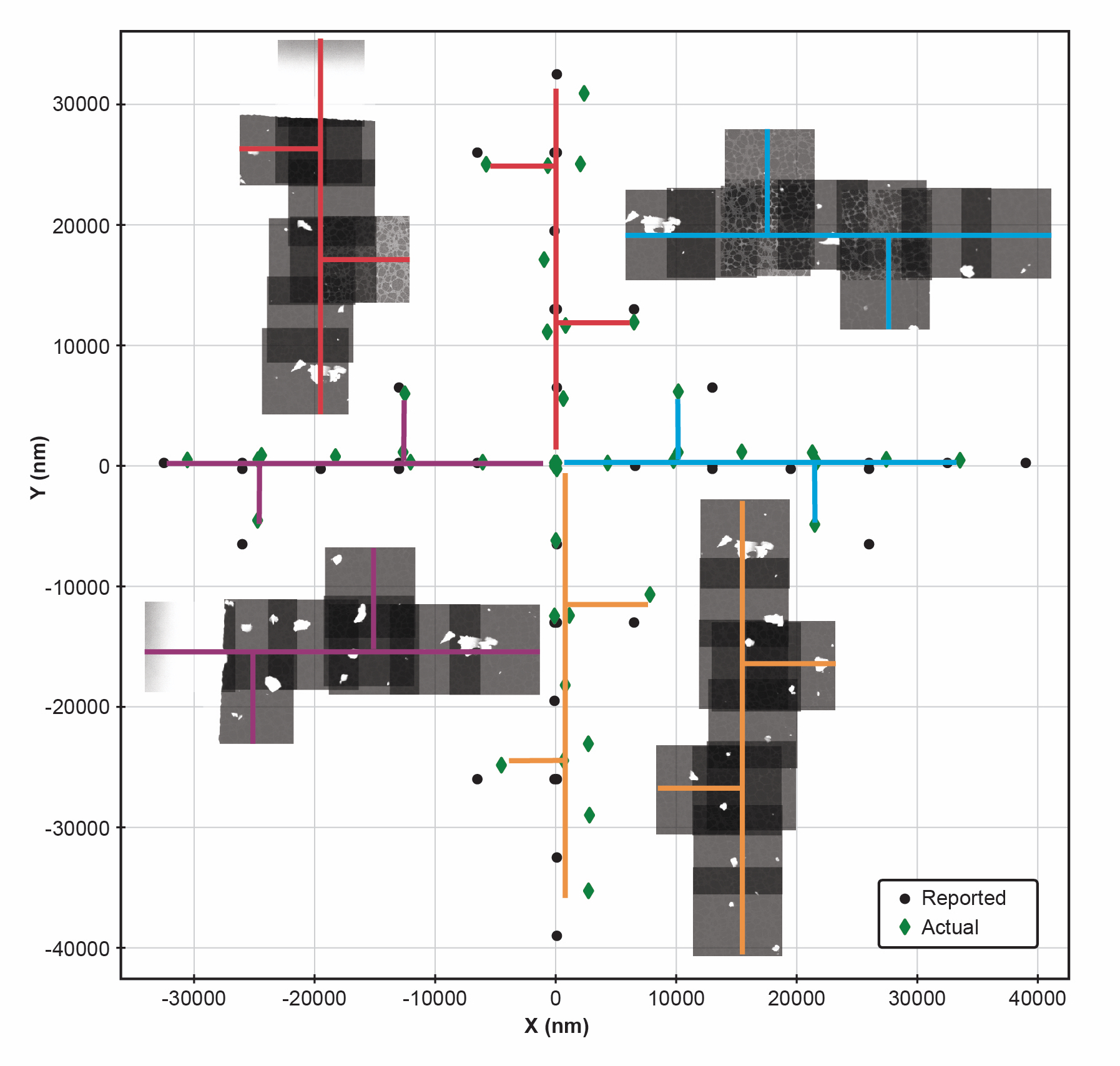}
\caption{\textbf{Hysteresis testing for coupled orthogonal axes.} Reported and actual locations for two-axis correlation acquisitions. Each cardinal direction moves from the origin outward and has the images overlapped at 50\% transparency to visualize the overlap. Each image was acquired at $50,000\times$ magnification with a field of view of 3.75 \textmu m by 3.75 \textmu m. \label{correlation}}
\end{figure}

Next, we check to see if movement along one axis impacts the measurements along a different axis, and vice versa. As shown in Figure \ref{stage_design}.b, movement within a TEM stage is characterized by arcs, not lines. Consequently, movement in an arc will cause displacements in more than one Cartesian axis, which if not accounted for by the microscope hardware or software, could couple movement commands between axes. For completeness, one should check every possible step direction and each possible combination of two. However, in this study we only consider the $x$ and $y$ directions, because these are the ones most easily interrogated for actual position location measurements using cross correlation techniques to align overlapping features. In this test, linear steps along a direction are interspersed with perpendicular steps as the microscope moves along each cardinal direction, as shown in Figure \ref{correlation}. In this figure, the reported and actual positions are plotted for movement in each cardinal direction to illustrate the discrepancies in the values.

In general, we observe that steps along different axes do not impact the reported locations on other ones. There is still hysteresis on each axis separately, so moving out and back does not return the microscope to the identical position, as can be seen with the inset microscope images. From the actual microscope images shown in Figure \ref{correlation}, it is apparent that the consistency of the reported position values is not reproduced in the actual images. This finding also highlights the fact that inaccuracies at the individual step level can accumulate over multiple move commands to create an overall larger error, where the final image is not the same as the initial image, even if the target or reported positions are identical. This finding indicates that orthogonal movements are largely uncoupled and we can compensate for lash independently. However, we observe that error compounds rapidly and that an object being tracked can move greatly within the field of view unless iterative correction is applied. To summarize the results of the hardware testing applied to our stage, we found that it shows hysteresis when changing directions and that the $x$ and $y$ axes are generally uncorrelated beyond this hysteresis. Due to the hysteresis observed in the first step in a given direction, there is a history dependence that makes general compensation of the motion to improve accuracy challenging. Within the general analysis framework shown in Figure \ref{alignment_cases}, our stage is in the category of Misalignment: Inconsistent shown in Figure \ref{alignment_cases}.f.

\clearpage

\section{Conclusions}

Here we have provided a general framework to evaluate stage movement any microscope. We show that there is coupling among axes of movement in the TEM, which operates in a spherical coordinate system that is much more complex than the Cartesian system commonly understood from SEM. We observe several potential complications for automated movement within this spherical frame of reference, owing to coupling of axes and imprecision in mechanical motion. Understanding this relationship is critical for emerging automated microscope systems and is particularly relevant for high-speed experimentation, where rapid movements must be made in sequence to track an object or reaction.

The analysis framework described here provides several concrete pieces of information relating to TEM stage motion. First, we found that in general, the target, reported, and actual position of the microscope might disagree with each other. Next, we identified that the first step in a given direction shows different behavior than repeated steps in that direction and this information can be used effectively in automation experiments. We observed that the $y$-axis tended to be more stable than the $x$-axis for our particular stage, indicating that it is important to map the characteristics of each holder. If there is a particular orientation where accuracy is more important in future experiments, for example, then this axis can be roughly aligned along the $y$-axis for improved performance as compared to a random alignment within the microscope. We emphasize that this analysis framework is general and our findings do not speak to overall relative performance of any specific holder design.

In general, we conclude that active feedback on the stage position in relation to regions of interest is essential to achieve automated human-level performance using present stages. This feedback may be provided through software correction, but it more desirable to eliminate imprecision at the hardware level. It is likely that full automation will be some combination of physical hardware and AI improvement. Currently, the reported stage position is merely inferred from the signal that is sent to the motors. This situation leaves an information gap within the system between the targeted and actual position, in addition to the misalignment reported for the system. Moving forward, we will wish to design new stages with precise, scalable movements, specifically incorporating independent axis control, accurate positioning feedback, as well as in situ stimuli needed for rich, automated experimentation. Such designs will no doubt be challenging to create, but they are an essential step toward breakthrough discoveries in the age of automation.

\clearpage

\section{Materials and Methods}\label{methods}

\subsection{Hardware}

The microscope used in this study is a customized probe-corrected JEOL GrandARM-300F STEM equipped with our AutoEM platform, that allows access and automation of low level controls such as magnifcation, tilt, and translation \citep{Olszta2021}. The data shown is acquired in STEM mode at 300 kV accelerating voltage at $50,000 \times$ magnification. Data processing is performed on a separate remote Dell Precision T5820 Workstation equipped with a Intel Xeon W-2102 2.9GHz processor and 1GB NVIDIA Quadro NVS 310 GPU. A JEOL low X-ray background double tilt holder was utilized in these experiments. The exact sequence of measurements for each calculation is given in Supplementary Information Section 1.

\subsection{Automated Data Collection}

The automation system is composed of interlinked hardware-software components, as described elsewhere \citep{Olszta2021}. HubEM acts as the main end-use application for the system. It serves as a point for entering configuration, storing data, and directing the cooperation of other components through inter-process communication. It is implemented in C\#/Python and uses Python.NET 2.5.0, a library that allows Python scripts to be called from within a .NET application. PyJEM Wrapper is an application that wraps the PyJEM 1.0.2 Python library, allowing communication to the TEMCenter control application from JEOL. It is written in Python and runs on the JEOL PC used to control the instrument. GMS Python allows communication to the Gatan Microscopy Suite (GMS) 3.4.3. It runs as a Python script in the GMS embedded scripting engine. All components communicate using a protocol based on ZeroMQ and implemented in PyZMQ 19.0.2.

\subsection{Image Registration}

Overlapping images were aligned using a custom Python 3.7.1 script that implements a normalized cross-correlation technique, as described in \citep{Olszta2021}. In brief, the images were converted to grayscale, normalized to have zero mean pixel intensity and maximum absolute pixel intensity of 1. The cross correlation was computed using the SciPy 1.7.3 library \citep{2020SciPy-NMeth}. Peak values in the cross correlation were used to determine optimal alignment and the actual positions reported in the text. 

\clearpage

\section{Acknowledgements}

The authors would like to thank Dr. Bethany E. Matthews for reviewing the manuscript. This work was supported by the Energy Storage Materials Initiative (ESMI) Laboratory Directed Research and Development (LDRD) program at Pacific Northwest National Laboratory (PNNL). PNNL is a multiprogram national laboratory operated for the U.S. Department of Energy (DOE) by Battelle Memorial Institute under Contract No. DE-AC05-76RL0-1830. The AutoEM system is located in the Radiological Microscopy Suite (RMS) in the Radiochemical Processing Laboratory (RPL) at PNNL.

\section{Competing Interests Statement}

The authors declare no competing interests.

\section{Author Contribution Statement}

K.R.F., M.J.O., D.H., S.A., and S.R.S. conceived and developed the research plan. K.Y. and M.J.O. conducted on-instrument testing. K.R.F., D.H., and C.D. programmed the testing routines. All authors contributed to the writing and editing of the manuscript.

\section{Data Availability Statement}

The image data shown and calculations of error are available on FigShare at: \url{https://doi.org/10.6084/m9.figshare.21735797}.

\clearpage

\bibliographystyle{MaM}
\bibliography{refs}

\clearpage

\renewcommand{\appendixname}{Supplementary Information}
\appendix*
\section{Testing Framework}

The main aim of this study is to construct a testing framework by which the motion of a microscope can be rigorously tested and evaluated. An important part of this goal is that the testing methods should be independent of the particular design or manufacturer of the microscope. For the tests described below, the only requirement is that the user be able to input target positions, record reported positions, and acquire images.  

The experimental investigation of the stage motion is broken into two broad categories. The first category is a set of consistency checks where the drift, jitter, repeatability of move commands, absolute location within the stage bounds, and magnification is changed are investigated. The second category is a set of state dependence checks where each axis is tested for hysteresis and the axes are checked for correlation between them. For each experiment, a target position is determined, the microscope stage is moved, an image is acquired and the reported microscope stage position is recorded. Using our AutoEM platform, this process can be automated for the rapid acquisition of data, but this capability is not a requirement.

\subsection{Consistency Checks}

To check the consistency of the stage, it is important to gauge how the stage is performing over time, both in terms of stability (drift and jitter) and then also in terms of repeatability. Time in this instance may be related to the number of movement commands issued, or it could relate to elapsed time while some external fluctuation is occurring (e.g. thermal equilibration or mechanical relaxation). Each of these experimental conditions is described in greater detail below.

\textbf{Drift:}
To measure the drift (change over time) of the stage, this experiment does not send move commands to the stage. Instead, one hundred images are acquired in exactly the same location, and the reported microscope position is also recorded each time an image is taken for purposes of comparison.

\textbf{Jitter:}
To measure the jitter (scatter of repeated commands) of the stage, this experiment repeats the structure of the drift experiment, but instead sends redundant move commands to the stage to move to exactly to where it is currently. The target position remains the same for each of the one hundred image acquisitions and can then be compared to the reported and actual positions.

\textbf{Repeatability:}
To test the repeatability of move commands, an initial location is selected. An image is acquired at this starting point, which functions as the origin for this particular experiment. Then, a move command is executed in a given cardinal direction (e.g. positive $x$) for a fixed step size and another image is acquired. The opposite move command is executed to (nominally) return to the origin and another image is acquired.  This process is repeated to build statistics, and then the whole procedure is repeated for the other cardinal directions ($+Y$, $-X$, and $-Y$). It is also possible to vary the step size over a range of scales. This approach allows the user to probe the dynamics of different movement mechanisms, which have different ranges of motion: for example, a small step size to test piezoelectric motion versus a large step size for testing stepper motor control. 

\textbf{Stage Location:}
To test whether the stage behaves similarly at multiple points within the possible parameter space of the stage position, any of the tests can be performed at multiple stage locations and compared. For example, testing could be done at the origin, near the maximum $x$ position, near the minimum $x$ position, the maximum $y$ position, and the minimum $y$ position, with the results checked for consistency, since an ideal stage should have similar behavior both in the middle of its travel range and at its extremes. This process could also be performed with a variety of tip and tilt conditions.

\textbf{Magnification:}
In a similar vein, another parameter to test for consistency is magnification. At a different magnification, the step size will necessarily be different to achieve feature overlap, so this can be effective tool to probe the regimes of different types of motors, such as the behavior of a stepper motor versus the behavior of a piezoelectric motor, when coupled with another test.

\subsection{State Dependence Checks}
Beyond consistency, another important factor when it comes to evaluating stage performance is whether the current state of the stage is dependent on the previous history of the stage. An ideal stage would have no memory of previous movements, but there are a variety of possible situations that could prevent this from being true. One practical example is backlash of a motor, where the current steps depend on the previous step. Another example would be nonlinear actuation of a piezoelectric. Steps near the limits of the motion have smaller travel distances than those that are near the middle. In certain cases, state dependence that is repeatable can be compensated for, but the following experiments can interrogate and quantify this behavior.

\textbf{Single Axis Hysteresis:}
To measure if there is hysteresis present on a movement axis, regular step commands of a fixed size in a single direction are issued. This process is repeated multiple times for a cardinal direction and then the direction of the step command is reversed an equal number of times to bring the target position back to the origin. This process is then repeated for the perpendicular cardinal directions.

\textbf{Two Axis Correlation:}
To measure if movement on one axis effects movement or stage positions on the other axis, an experiment where movement in a given direction is programmed. After two steps in the cardinal direction, a step in the perpendicular cardinal direction is performed with an image acquisition, followed by an immediate backtrack of this step and another image acquisition. Two more steps in the original cardinal direction occur and then a step occurs in the perpendicular cardinal direction, starting the opposite way as the previous perpendicular step occurred. As before, the perpendicular step is undone and take final steps are taken in the original cardinal direction. This process is repeated for the other three possible cardinal directions.

\section{Bounds on Expected Stage Performance}
We used a variety of the preceding tests to evaluate the performance of our stage, in addition to the measurements described in the main text. Here we summarize the overall findings to provide bounds on the expected stage performance, both in terms of the reported position and the actual position.

\subsection{Target Position Vs. Reported Position}
\begin{figure}
\includegraphics[width=\textwidth]{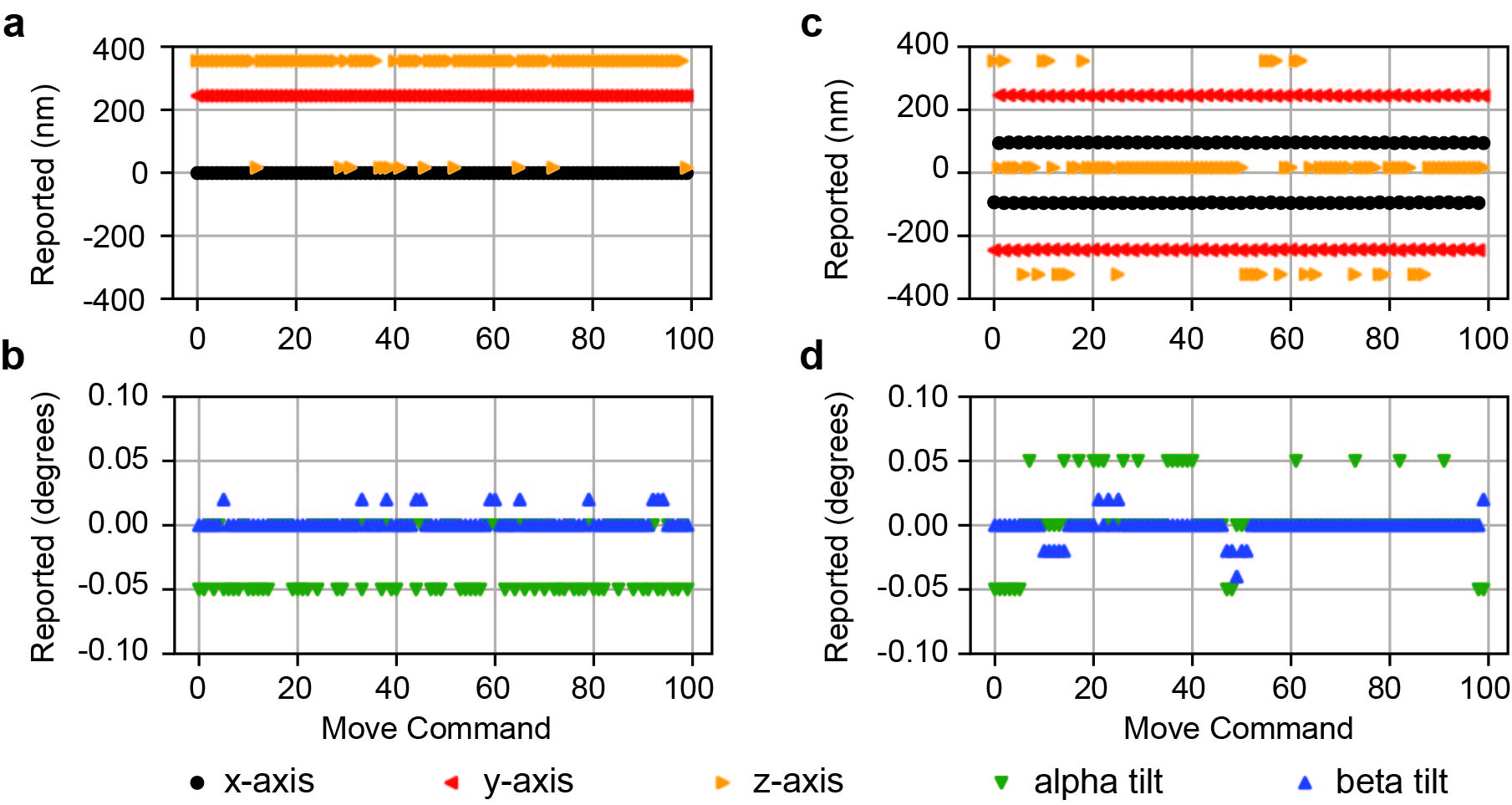}
\caption{\textbf{Single axis movement.} (a) Reported position as a function of move command for the translational axes ($x$, $y$, and $z$) measuring drift. (b) Reported position as a function of move command for the rotational axes ($\alpha$ and $\beta$) measuring drift. (c) Reported position as a function of move command for the translational axes ($x$, $y$, and $z$) measuring jitter. (d) Reported position as a function of move command for the rotational axes ($\alpha$ and $\beta$) measuring jitter.  \label{target_reported_driftjitter}}
\end{figure}

For the drift experiments described previously, the microscope was moved to a location, all coordinates ($x$, $y$, $z$, $\alpha$, $\beta$) set equal to 0, and then the microscope reported its position without moving one hundred times. An ideal stage would have no change over time, with all reported values equal to 0. As shown in Figure \ref{target_reported_driftjitter}.a--b, the $x$ and $y$ values were consistent over time, with the $x$ values being very close to 0. The reported values for the $z$ axis, the $\alpha$ axis, and the $\beta$ axis all showed changes over time, so this forms a lower limit on the possible accuracy for reported movements on these axes. This sort of measurement is a measurement of the statistical uncertainty in any reported value from the microscope. For $z$, this is 349 nm spread in measurements. For $\alpha$, the spread is $0.02^{\circ}$ and for $\beta$ it is $0.05^{\circ}$.

For the jitter experiments, the microscope was kept at the same location, but now it was told to move to that identical location one hundred times, with the reported values being recorded each time. As seen in Figure \ref{target_reported_driftjitter}c--d, if the microscope has been told to go to a specific location previously and is told to go there again, then it will report a potentially different value each time. The $x$-axis and $y$-axis values flip approximately symmetrically about the desired target location (-94 to 96 nm for $x$, and -244 to 244 nm for $y$). The $z$ axis, the $\alpha$ axis, and the $\beta$ axis all show more erratic behavior, consistent with the statistical limits that were observed in the drift experiments in the left half of the figure.  

\textbf{Bounds on Reported Stage Positions:}

In summary, for this particular stage/holder, we now have bounds on the performance that can be expected for the reported positions. For the $x$-axis, we would not expect accuracy to the target position to within $\pm100$ nm. For the $y$-axis accuracy is bounded by $\pm 250$ nm and for the $z$-axis it is bounded by $\pm 350$ nm. In terms of the rotational axes, $\alpha$ will not be accurate below $\pm 0.05^{\circ}$ and $\beta$ will not be accurate within $\pm 0.02^{\circ}$. On top of these limits, the first step in a given direction will be larger than subsequent steps, as shown by the hysteresis experiments in the state dependence checks. Typically, the size of the difference for the first step is within the bounds stated above for the accuracy. However, this effect is distinctly noticeable when performing repeated motions, as would occur during the acquisition of a large area montage.

\subsection{Target Position Vs. Actual Position}
\begin{figure}
\includegraphics[width=\textwidth]{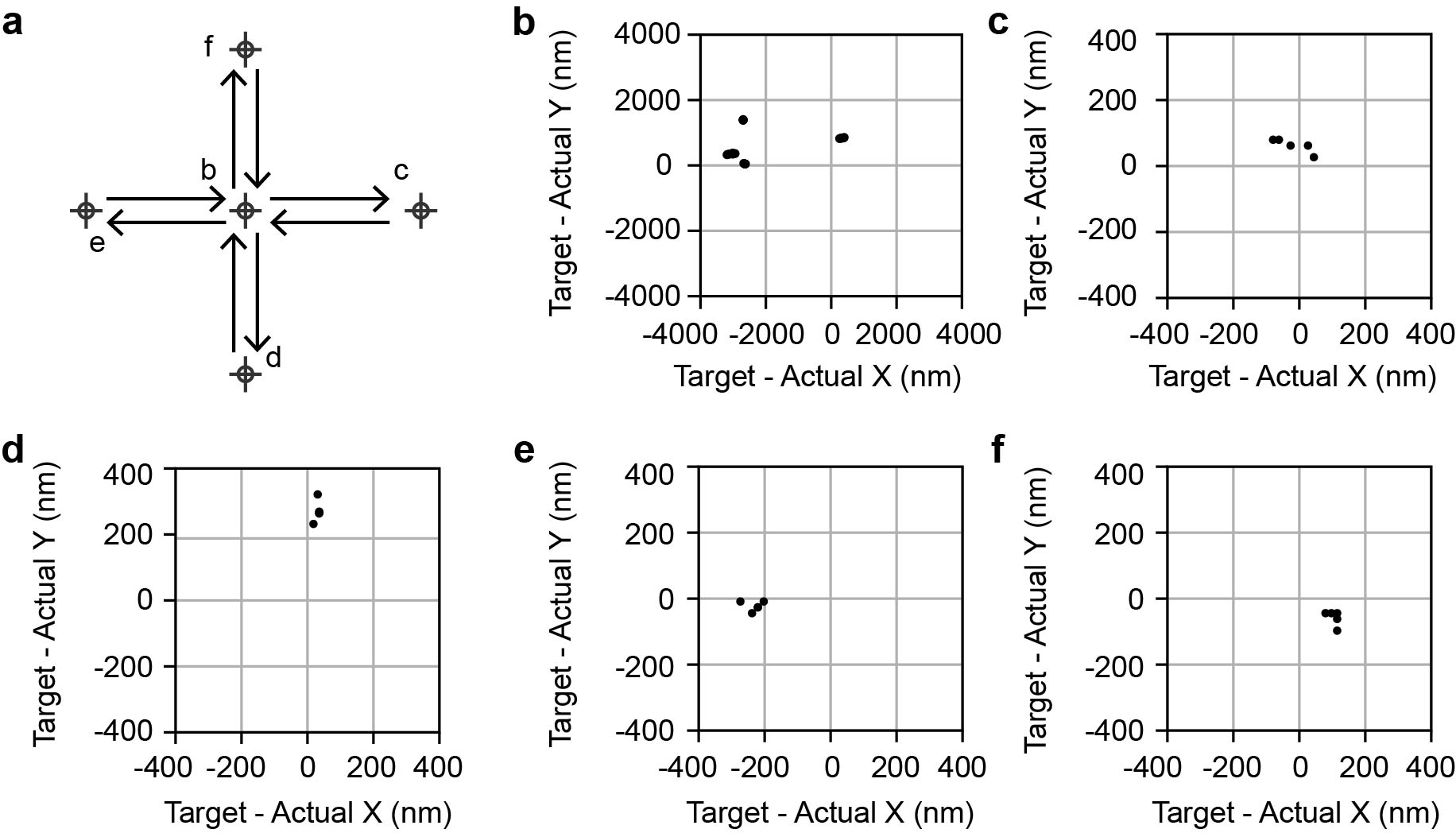}
\caption{\textbf{Repeatability.} (a) Illustration of the locations used for testing the repeatability of movements in the cardinal directions. (b) Actual position displacements for the center point, (0,0). (c) Actual position displacements for steps in the positive $x$-direction. (d) Actual position displacements for steps in the negative $y$-direction. (e) Actual position displacements for steps in the negative $x$-direction. (f) Actual position displacements for steps in the positive $y$-direction. \label{target_actual_repeatability}}
\end{figure}

One limitation of the overlap technique used for image alignment is that it is only possible to compare images which have identifiable features that are identical across two images. As such, we have broken up the repeatability check into a set of smaller comparisons. For each movement type, we can compare the relative position to the first in the series. It also restricts us to providing bounds on the actual position to only the $x$ and $y$ coordinates. This provides us with a set of five distinct locations for analysis, as seen in Figure \ref{target_actual_repeatability}. A schematic of the locations is shown in Figure \ref{target_actual_repeatability}.a, where we start in the center and step around the cardinal directions in a clockwise fashion starting with positive x-direction steps. The ideal situation for these results would be for the data to be scattered symmetrically about (0,0) in each case. If this situation cannot occur, then it would be next best for them to be off by a consistent amount for each different direction. Unfortunately, for this stage/holder, that does not appear to be the case. While all of the individual step directions, Figure \ref{target_actual_repeatability}.c--e, show approximately the same magnitude of being within 400 nm in $x$ and $y$, they do not show a consistent pattern to their displacements. Furthermore, the results for the center, Figure \ref{target_actual_repeatability}.b, are dramatically larger, with the required bounds being 3000 nm, primarily as a result of a large $x$ displacement.  

\textbf{Bounds on Actual Stage Positions:}
As was done for the reported stage positions, for this particular stage/holder, we now have bounds on the performance that can be expected on the actual position of the $x$- and $y$-axes. For the $x$-axis, we would not expect accuracy relative to the target position to within $\pm3000$ nm in the worst case. A more typical bound would be approximately $\pm350$ nm, as seen in the repeatability experiments. The performance for the $y$-axis was somewhat better, as here accuracy is bounded by $\pm2000$ nm in the worst case and by $\pm350$ nm in a more typical case. We note that these experiments cannot interrogate the $z$-axis or the rotational axes. On top of these statistical limits, the first step in a given direction will be smaller than subsequent steps, which is different than the behavior which was observed in the reported positions.  

\subsection{Summary of Expected Stage Performance}

To summarize the results of the experiments to test the reported and actual positions, the data from the previous sections has been put into Table \ref{table:bounds}.  Generally, the reported position tends to agree fairly closely with the target position. The worst case bounds on the actual position are much larger due to the hysteritic effect of changing movement directions. The values reported are worst case values, but with careful experiment design it is possible to reduce these by minimizing (e.g. by the number of direction changes or by overshooting the target position and always approaching from a consistent direction).

\begin{table}
\begin{center}
\begin{tabular}{ c | c | c }
 Axis       & Reported Position      & Actual Position \\ 
 \hline
 x          & $\pm100$ nm         & $\pm3000$ nm \\  
 y          & $\pm250$ nm         & $\pm2000$ nm \\
 z          & $\pm350$ nm         & N/A \\  
 $\alpha$   & $\pm0.05^{\circ}$   & N/A  \\
 $\beta$    & $\pm0.02^{\circ}$   & N/A 
\end{tabular} 
\caption{Statistical bounds on the worst case values for the reported and actual positions as evaluated by the analysis framework.}
\label{table:bounds}
\end{center}
\end{table}

\end{document}